# DETAILED ANALYSIS OF THE RESPONSE OF DIFFERENT CELL LINES TO CARBON IRRADIATION

Hana Hromčíková[*], Pavel Kundrát, Miloš Lokajíček
Institute of Physics, Academy of Sciences of the Czech Republic, Na Slovance 2, 18221 Praha 8, Czech Republic

**Published survival data for Chinese hamster ovarian cells CHO-K1 and their radiosensitive mutant xrs5 after irradiation by carbon ions of energies from 2.4 to 266.4 MeV/u have been analyzed using the probabilistic two-stage radiobiological model, which enables to represent the interplay of damage induction and repair processes. The results give support for the hypothesis that the differences in radiation sensitivity of diverse cell lines are given primarily by their different repair capabilities, and indicate the need for explicitly representing the outcome of repair processes in radiobiological models and treatment planning approaches in radiotherapy.**

## INTRODUCTION

Differences in damage induction, especially in the yields of complex double-strand breaks (DSBs) and other types of clustered lesions[1], have been shown to correlate with the differences in the response to radiations of diverse quality. On the other hand, the differences in the response to the same radiation kind observed among different cell lines seem to be given primarily by their different repair capacities. To test this hypothesis, survival data for two hamster cell lines irradiated by carbon beams of different energies[2] have been analyzed in detail. The studied cell lines are the wild-type Chinese hamster ovarian line CHO-K1 and its radiosensitive mutant xrs5, which differs from the wild-type line by lacking the Ku80 subunit of the active DNA-PK complex involved in the DSB repair. Accordingly, the two lines were found to differ significantly in the probabilities of repairing the damage successfully, while it has been possible to neglect minor differences in damage induction.

## RADIOBIOLOGICAL MODEL

To represent the damage induction and repair processes, the probabilistic two-stage model[3] has been used. In the model, the surviving fraction at dose $D$ is given by

$$s(D) = \Sigma_k P_k(D)\, q_k \,, \qquad (1)$$

where survival probability of cells traversed by $k$ tracks

$$q_k = [1-(1-(1-a)^k)(1-r_k^a)][1-(1-(1-b^2)^{k(k-1)/2})(1-r_k^b)] \,. \quad (2)$$

The distribution of track numbers $k$ over cell nuclei is described by Poisson statistics $P_k$ with the mean number proportional to geometrical cross-section of the nucleus. Average probability that a single track induces lethal damage to DNA has been denoted by $a$. Similarly, $b$ stands for the probability that a less severe lesion has been formed, which has to combine with at least another one to be lethal ("combined lesions")[3]; the term $(1-b^2)^{k(k-1)}$ approximates the combined effects of $k$ tracks[3,4].

## ANALYSIS OF EXPERIMENTAL DATA

Survival data[2] for CHO-K1 and xrs5 cells irradiated by carbon ions with energies from 2.4 to 266.4 MeV/u (LET 13.7 – 482.7 keV/µm) has been analyzed by the given model. Damage probabilities per track in dependence on LET $L$ have been parameterized by

$$a(L) = \frac{a_0\left(1-\exp\left(-(a_1 L)^{a_2}\right)\right)}{1+a_3\exp\left(-(a_1 L)^{a_2}\right)},$$

$$b(L) = \frac{b_0\left(1-\exp\left(-(b_1 L)^{b_2}\right)\right)}{1+b_3\exp\left(-(b_1 L)^{b_2}\right)} \qquad (2)$$

identically in both cell lines. For the sake of simplicity, only lesions that are not repaired by xrs5 cells have been considered in the analysis. Hence $r = 0$ for xrs5 cells, while for CHO-K1 cells the repair success probability has been parameterized by

$$r^a(L,k) = \frac{r_0\left(1-\exp\left(-(r_1 L)^{r_2}\right)\right)}{1+r_3\exp\left(-(r_1 L)^{r_2}\right)} \,; \qquad (3)$$

repair of combined lesions has been neglected, $r^b = 0$. Geometrical cross sections of cell nuclei reported in the given experiment[2] were $\sigma_{CHO-K1} = 108$ µm$^2$ and $\sigma_{xrs5} = 122$ µm$^2$, respectively.

Model representation of survival curves for both cell lines irradiated at different beam energies are presented in Figure 1. The derived damage induction and repair success probabilities are shown in Figures 2 and 3.

*Corresponding author: Hromcikova@seznam.cz

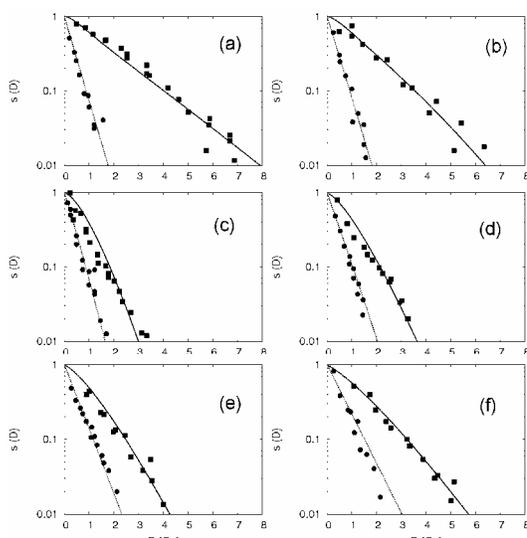

Figure 1: Model representation (lines) compared to experimentally measured survival (points) for CHO-K1 (squares) and xrs5 cells (circles) irradiated by mono-energetic carbon beams of 13.7 (panel a), 32.4 (b), 153.5 (c), 275.1 (d), 339.1 (e) and 482.7 keV/μm (panel f), respectively.

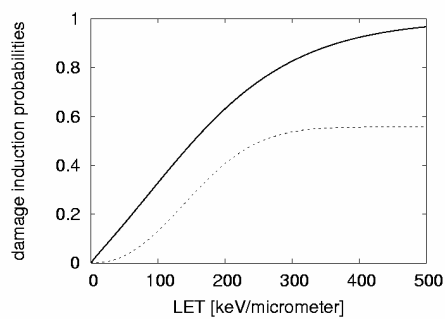

Figure 2: LET-dependent probabilities of carbon tracks to induce single-track (solid line) and combined lesions (dashed line).

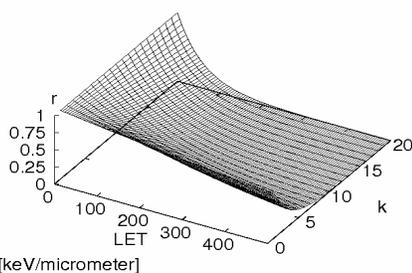

Figure 3: Repair success probability for single-track lesions in dependence on the number of tracks $k$ and LET.

## DISCUSSION AND CONCLUSION

While the differences between the two cell lines in damage induction by single carbon tracks might have been neglected, the derived repair success probability (Figure 3) demonstrates that the wild-type CHO-K1 cells succeed in repairing large amounts of damage not repaired by xrs5 cells. This finding directly reflects the reduced DSB repair capacity of xrs5 cells.

The present results indicate that differences in repair capacities among diverse cell lines account for the differences in their intrinsic radiation sensitivity. In simplified model schemes[4-6], only the effective damage yields (yields of lethal damage not repaired by the cell) have been considered, i.e. the effects of repair processes have been implicitly included in the description of damage induction. Although such simplified approaches can be used successfully in analyzing the differences in response to diverse radiation qualities, their drawback is that the intrinsic radiation sensitivity of different cell lines is only approximated. The present results stress the need for further developing detailed biology-oriented models of radiobiological effects, which should reflect the results of damage induction and repair studies.

*Acknowledgment:* This work was supported by the grant "Modelling of radiobiological mechanism of protons and light ions in cells and tissues" (Czech Science Foundation, GACR 202/05/2728).